# LIVE-MARKER: A PERSONALIZED WEB PAGE CONTENT MARKING TOOL

K S Kuppusamy[1] & G.Aghila[2]

The tremendous amount of increase in the quantity of information resources available on the web has made the total time that the user spends on a single page very minimal. Users revisiting the same page would be able to fetch the required information much faster if the information that they consumed during the previous visit(s) gets presented to them with a special style. This paper proposes a model which empowers the users to mark the content interesting to them, so that it can be identified easily during successive visits. In addition to the explicit marking by the users, the model facilitates implicit marking based on the user preferences. The prototype implementation based on proposed model validates the model's efficiency.

## 1. Introduction

The colossal size of World Wide Web has constructed a huge arena of choices to the user when he/she is looking for information. This is evident from the fact that search engines return a long list of results for any query. This information overload problem has made the user to spend very little time on average on any page.

There exist two different aspects which need to be considered when users are visiting web pages. Different users visiting the same page would be interested in different portions of that page. The portions which are interesting to the user are decided by their profile. When user visits the page for the successive times, he/she has no clue on what information was useful in that page, during the previous visit(s). If the information interesting to the user is given a specific style then it would make them easier to locate the necessary information.

This research work is carried out considering the above specified two factors. The objectives of this research work are as listed below:

- Proposing the "LiveMarker" model to provide personalized content marking for the web pages.

- Validating the model empirically by conducting experiments with users with different set of profile-keywords.

The rest of this paper is organized as follows. Section 2 would highlight the works related to the proposed model. Section 3 illustrates the proposed "Live-Marker" model and provides algorithms. Section 4 is about experimental setup. Section 5 list outs the conclusions and future directions of this research work.

## 2. Related Works

Personalization is one of the active research topics in the web based information retrieval domain. Personalization is the process of customizing based on the user requirements and preferences. The work presented in [1], proposes a method which utilizes the experiences of the earlier users in a collaborative manner. Generally, the personalized result rendering is based upon the "feedback" from the end-users.

There exist two types of feedbacks: Implicit and Explicit. In the explicit feedback mechanism user has to explicitly indicate the relevant and non-relevant items. In the case of implicit feedback it would gathered automatically based on the actions performed by the user. Here the user is not required to explicitly mark it as relevant or not relevant. Both these types of feedbacks are discussed in [2], [3] [4]. The method used in [5] provides an approach to highlight text in the browser window. The proposed model incorporates both implicit and explicit feedback methods.

## 3. The Model

The proposed model requires the user to create their profiles. The profile would consist of various keywords that depict the interest of the user. These profile-keywords are as shown in (1).

$$\Gamma = \{\mu_1, \mu_2 \ldots \mu_n\} \quad (1)$$

Let us denote the page that the user visits as P. This page would consist of set of terms as shown in (2).

$$P = \{\varepsilon_1, \varepsilon_2 \ldots \varepsilon_m\} \quad (2)$$

[1]Department of Computer Science, Pondicherry University, Pondicherry, INDIA
[2]Department of Computer Science, Pondicherry University, Pondicherry, INDIA
E-mail: [1]kskuppu@gmail.com, [2]aghilaa@yahoo.com

Two types of markings are there in the proposed model. One is the explicit marking by the user ($\Omega$) and another is the implicit marking ($\Psi$) made by the system based on the profile-keywords. The Marker component is represented as shown in (3)

$$M = \langle \Omega, \Psi \rangle \qquad (3)$$

When the user visits the page for the first time there won't be any explicit marking. The implicit markings on the page are as shown in (4).

$$\Psi = \{P \cap \Gamma\} \qquad (4)$$

The original source page is now updated for display in the client side which is denoted as $\overline{P}$. The modified page with initial implicit markings is as shown in (5).

$$\overline{P} = \begin{bmatrix} \Psi & \forall (\varepsilon_i \in P, \mu_i \in \Gamma) : \overline{P} \cup (\varepsilon_i \cap \mu_i) \\ \Omega & \varnothing \end{bmatrix} \qquad (5)$$

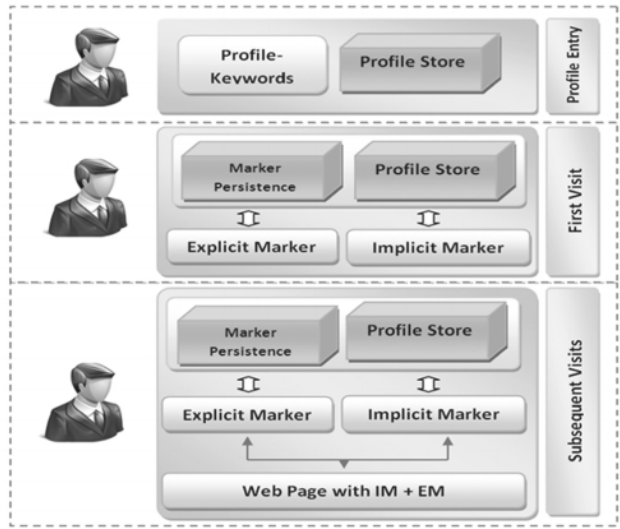

Fig. 1: The Livemarker Model

In (5) the explicit marking component $\Omega$ is having the value $\varnothing$ because there won't be any explicit marking during the first visit. In the successive visits the user can mark the contents according to his/her needs. The page with both explicit and implicit markings is as represented in (6).

$$\overline{P} = \begin{bmatrix} \Psi & \forall (\varepsilon_i \in P, \mu_i \in \Gamma) : \overline{P} \cup (\varepsilon_i \cap \mu_i) \\ \Omega & \bigcup_{i=1}^{k} \varepsilon_i \end{bmatrix} \qquad (6)$$

The value of 'k' shown in (6) indicates the number of terms that the user has marked explicitly.

Hence the user can mark only the content available on the page the explicit marking set is always a subset set of P. The same condition is true for implicit markings also as shown in (7).

$$\Psi \subseteq P; \Gamma \subseteq P \qquad (7)$$

The architecture of the "Live-Marker" model is illustrated in Fig.1. The three different blocks in Fig.1. illustrate the user's profile creation stage, first visit and successive visits.

Algorithm LiveMark

Input: Page P

Output : Marked Page $\overline{P}$

Begin

    //Fetch the contents of the page P.

    P = getContents(url)

    If (firstVisit(P))

    Begin

        //Fetch the profile_keywords

        pk = fetch_keywords();

        for each term in page P

        highlight the intersecting terms in P and pk

    End

    Fetch the highlight data from mark_ persistency (mp)

    Highlight the terms intersecting in Page P and mp

End

## 4. EXPERIMENTS AND RESULTS ANALYSIS

In order to empirically validate the model a prototype implementation was done as a FireFox extension. The decision to go for the "extension" based approach is motivated by the fact that it gets tightly integrated with the browser. This makes the user free from depending on any another tool apart from the browser. The screenshots of profile keywords entry is given in Fig.2. The Fig. 2 (a) shows the screenshot of profile-keywords in the SQLite manager extension [6], [7]. Fig. 2 (b) is a sample marked web page. The terms highlighted in green color refers to implicit markings. The yellow color highlighting refers to explicit markings. The Table 1 lists the explicit and implicit markings count for three different users. It can be observed from the table that different users visiting the same page are getting different terms highlighted and their count also differs. The Fig.3 depicts the comparative charts of Implicit and Explicit markings for three different users.

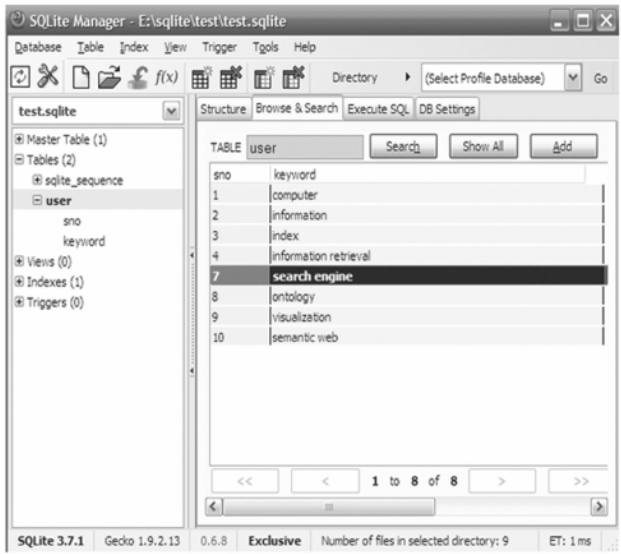 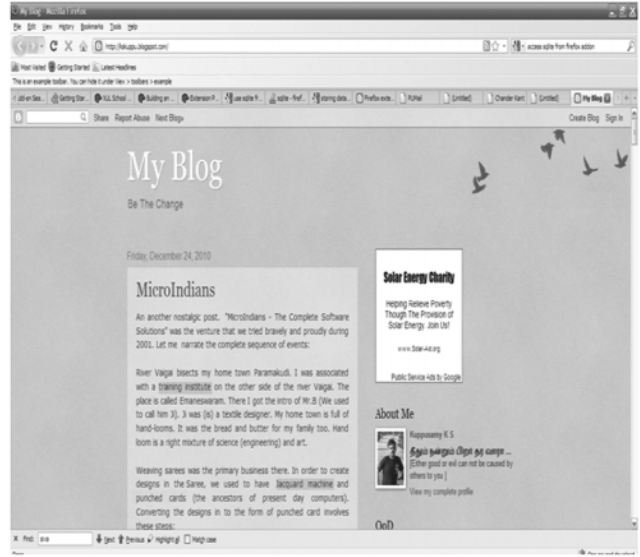

(a)                                                  (b)

Fig.2. (a) The Profile Keywords Through SQLite Manager (b) The Highlighted Page

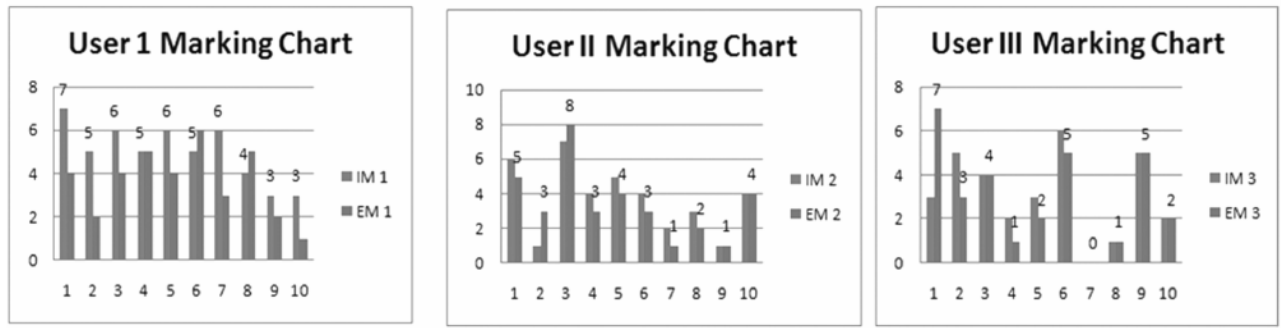

Fig.3: The Marking Chart for Three Different Users for a Set of 10 pages

### 5. CONCLUSION AND FUTURE DIRECTIONS

The proposed model enables the user to locate the information specific to him/her in a page more easily. The two different styles of markings help the user to distinguish between the implicit markings made by the system and the explicit markings made by the user.

The future directions for this research work include the following: Extending the profile-keywords based approach to an Ontology based approach; incorporating additional components like images for marking in addition to the text marking proposed in the current model.

#### Table 1
#### Terms Highlighting

| Page ID | User I | | User II | | User III | |
|---|---|---|---|---|---|---|
| | IM | EM | IM | EM | IM | EM |
| 1 | 7 | 4 | 6 | 5 | 3 | 7 |
| 2 | 5 | 2 | 1 | 3 | 5 | 3 |
| 3 | 6 | 4 | 7 | 8 | 4 | 4 |
| 4 | 5 | 5 | 4 | 3 | 2 | 1 |
| 5 | 6 | 4 | 5 | 4 | 3 | 2 |
| 6 | 5 | 6 | 4 | 3 | 6 | 5 |
| 7 | 6 | 3 | 2 | 1 | 0 | 0 |
| 8 | 4 | 5 | 3 | 2 | 1 | 1 |
| 9 | 3 | 2 | 1 | 1 | 5 | 5 |
| 10 | 3 | 1 | 4 | 4 | 2 | 2 |